\documentclass[
 aps, pra,
 amsmath,amssymb,
 11pt,
 final,
tightenlines,
 twoside,
 twocolumn,
 nofloats,
nofootinbib,
 superscriptaddress,
showkeys,
showkeywords,
 ]
{revtex4-2}

\usepackage[T1]{fontenc}
\usepackage[utf8x]{inputenc}
\usepackage[english]{babel}
\usepackage{graphicx}% Include figure files
\usepackage{dcolumn}% Align table columns on decimal point
\usepackage{bm}% bold math

\input{maik.rty}

\setcitestyle{authoryear,round}
\def\squareforqed{\hbox{\rlap{$\sqcap$}$\sqcup$}}

\def\sq{\ifmmode\squareforqed\else{\unskip\nobreak\hfil
\penalty50\hskip1em\null\nobreak\hfil\squareforqed
\parfillskip=0pt\finalhyphendemerits=0\endgraf}\fi}

\def\utw{\smash{\rlap{\lower5pt\hbox{$\sim$}}}}

\def\udtw{\smash{\rlap{\lower6pt\hbox{$\approx$}}}}

\def\diameter{{\ifmmode\mathchoice
{\ooalign{\hfil\hbox{$\displaystyle/$}\hfil\crcr
{\hbox{$\displaystyle\mathchar"20D$}}}}
{\ooalign{\hfil\hbox{$\textstyle/$}\hfil\crcr
{\hbox{$\textstyle\mathchar"20D$}}}}
{\ooalign{\hfil\hbox{$\scriptstyle/$}\hfil\crcr
{\hbox{$\scriptstyle\mathchar"20D$}}}}
{\ooalign{\hfil\hbox{$\scriptscriptstyle/$}\hfil\crcr
{\hbox{$\scriptscriptstyle\mathchar"20D$}}}}
\else{\ooalign{\hfil/\hfil\crcr\mathhexbox20D}}%
\fi}}

\begin{document}

\selectlanguage{english}


%\ydk{}
%\titlerunning{}
%\authorrunning{}
%\toctitle{}
%\tocauthor{}

\title{Indicatives of Early Stages of Star Formation in the Universe}

\author{\firstname{I.~A.}~\surname{Acharova}%\href{URL}{https://orcid.org/0000-0002-4942-6905}
}%https://orcid.org/0000-0002-4942-6905
 \email{iaacharova@sfedu.ru}
 \affiliation{Southern Federal University, Rostov-on-Don, 344006 Russia}
%\href{URL}{https://orcid.org/0000-0002-4942-6905}
\author{\firstname{M.~E.}~\surname{Sharina}} %\href{URL}{https://orcid.org/0000-0001-9840-5580}
 \affiliation{Special Astrophysical Observatory,  Russian Academy of Sciences,  Nizhnii Arkhyz, 369167 Russia}

%\received{July 18, 2025}  \revised{September 22, 2025} \accepted{October 13, 2025}

\begin{abstract}
The paper analyzes formation conditions for globular clusters (GCs) in circumgalactic clouds. The similarity between the metallicity distributions of GCs in the nearby Universe and of circumgalactic clouds is substantiated in detail over a wide range of redshifts: from \mbox{0.2} to \mbox{5.9}. The distributions of the number of circumgalactic clouds and GCs both contain a sequence of four local maxima at the metallicity values: \mbox{$[\rm{X/H}]\simeq -2.6,  -2.0,  -1.4,-0.5$}. The sequential enrichment of a circumgalactic cloud with a mass of $10^{8}\,M_{\odot}$ is calculated starting the extremely low metallicity \mbox{$ [\rm{X/H}] <-2.3$}, then following through the stages of \mbox{$-2.3 \le [\rm{X/H}]<-1.7$} and \mbox{$-1.7 \le [\rm{X/H}] < -0.9$} to the high metallicity \mbox{$[\rm{X/H}] \ge -0.9$}, where the boundaries of these ranges coincide with the local minima of the number of objects in the distributions. It is shown that for the reproduction of such distributions, it is sufficient that at each stage of enrichment of a part of a cloud in metals, one or more GCs with a total mass of \mbox{$3 \times 10^{6}\,M_{\odot}$} are formed. It is shown that the maximum mass of stars capable of leading to supernova explosions increases with the increase of metallicity. Possible values of this mass are calculated for the metallicities corresponding to the maxima in the distributions of clouds and GCs.
\end{abstract}

\maketitle

\section{INTRODUCTION}
Analysis of the metallicity distribution of circumgalactic clouds within a wide range of redshifts, \mbox{0.2--5.9}, and globular clusters in the nearby Universe allows us to clarify the processes having occurred at the initial stages of galaxy formation. Since the concepts of ``iron abundance'' and ``metallicity'' will be used in the text, we note that $[\rm{Fe/H}]$ denotes the iron abundance. Metallicity will be denoted with $[\rm{X/H}]$, where \mbox{$[\rm{X/H}]=[\alpha/\rm{H}]$} or \mbox{$[\rm{X/H}]=[\rm{Fe/H}] + \Delta$}, in this case $\Delta$ takes different values, \mbox{from 0.3} \mbox{to 0.5} depending on the element, on the basis of which the metallicity is determined (Pritzl et al., 2005; Lehner et al., 2019; {Nu{\~n}ez} et al., 2022).

As a lower limit of metallicity, at which the formation of low-mass stars occurs, it is natural to use the observed lower limit of metallicity of the stars in the Galactic halo, which falls at \mbox{$[\rm{X/H}]\sim{-3.5}$\,dex} and corresponds to \mbox{$[\rm{Fe/H}]\sim{-4.0}$\,dex} (Cohen et al., 2013; Roederer and Kirby, 2014). This limit is consistent with the values obtained in Wise and Abel (2008) and Chiaki and Wise (2019) showing that a supernova originating from a Population~III star can enrich gas with a mass of \mbox{$10^5$--$10^6\,M_{\odot}$} in metals to an abundance of \mbox{$10^{-4}\,Z_{\odot}$ ($Z_{\odot}$} is the solar metallicity), if it has been an explosion of a single star with a mass of up to $130\,M_{\odot}$; and up to an abundance of $10^{-3}\,Z_{\odot}$, if the supernova explosion has occurred as a result of a pair-unstable mechanism (possible for stars with masses from $130\,M_{\odot}$ to $250\,M_{\odot}$, when the gamma radiation arising from the collapse of their core is capable of producing electron-positron pairs) (Heger et al., 2003). As is noted in the paper by Shchekinov et al. (2023), incomplete mixing of metals in the enriched gas results in the oldest low-mass stars having a significant spread in metallicity, more than an order of magnitude.

The lower limit on the stellar metallicity found in the above-cited papers is in good agreement with the lowest metallicity of stars in all studied ultra-faint dwarf galaxies (UFDs) (Simon, 2019; Fu et al., 2023) near \mbox{$[\rm{Fe/H}]\sim-4.0$\,dex}, but is noticeably lower (by approximately $0.7$\,dex) than the lowest observed metallicity of GCs in the nearby Universe (Beasley et al., 2019).

The formation of GCs and UFDs begins separately from each other in different regions of space at the same time, about 13\;Gyr ago, taking into account the age determination error (Chattopadhyay et al., 2012). Their properties are very different. UFDs are objects with very low stellar density and a predominance of the dark matter, while GCs are, in most cases, dense stellar conglomerates without the dark matter.
The metallicity of GCs and circumgalactic clouds is multimodal, i.e., it shows maxima and minima in the distribution of the number of objects by metallicity (Harris, 2010; Shapiro et al., 2010; Kruijssen, 2015; Gratton et al., 2019). According to the estimates presented in this paper (see Section\,2), local maxima occur at \mbox{$[\rm{X/H}] \simeq -2.6$}, \mbox{$-2.0$}, \mbox{$-1.4$}, $-0.5$, and local minima at \mbox{$ [\rm{X/H}] \simeq -2.3$}, \mbox{$-1.7$}, $-1.0$. Taking into account the masses of circumgalactic clouds \mbox{($10^8$--$10^9\,M_{\odot}$)} and GCs (approximately $10^6\,M_{\odot}$), this distribution is indicative of three stages of rapid, compared to the star-formation characteristic time, heavy-element enrichment, which is possible only as a result of the coherent explosion of hundreds of thousands of supernovae (see Section\;4).

In this paper, we study the conditions of star formation in circumgalactic clouds at the GC formation stage. Conclusions are based on an analysis of the successive enrichment of clouds in alpha elements, the main sources of which are supernovae originating from massive (with a mass greater than $8\,M_{\odot}$) stellar progenitors---core-collapse supernovae (Woosley and Weaver, 1995; Smartt, 2009).

Cloud masses in different metallicity ranges can be considered proportional to the frequency of occurrence of clouds of the same density in these ranges. For example, if the frequency of occurrence is the same, then the characteristic size of clouds in different metallicity ranges is also the same. This is because the line of sight to the quasar intersects the clouds randomly. Therefore, in this case, the masses of the enriched and unenriched parts of the cloud are equal to a first approximation.

\section{ANALYSIS OF THE PROPERTIES OF CIRCUMGALACTIC CLOUDS}

In the paper by Acharova et al. (2022), it was established that circumgalactic clouds up to the redshift $z=5$ are likely to be the remnants of the parent clouds, in which GCs were formed. Let us briefly recall the main premises of this conclusion. Circumgalactic clouds and GCs of our and other galaxies in the region with \mbox{$[\rm{X/H}] \ge -1.7$} show a division into subgroups with high and low metallicity, with two clearly defined maxima in the distribution of the cloud number from metallicity at \mbox{$ \langle [\rm{X/H}] \rangle =-0.5 \pm 0.3$} and \mbox{$ \langle [\rm{X/H}] \rangle =-1.4 \pm 0.3$} and a deficiency of clouds near \mbox{$[\rm{X/H}]\sim-0.9$}. The 15-fold enrichment in metals occurred rapidly: the ages of GCs in both groups are identical (within measurement error). As shown in the studies by Acharova and Sharina (2018) and Acharova et al. (2022), such a sharp and rapid enrichment can only be explained by a coherent explosion of hundreds of thousands of supernovae, that is, occurring over a very short time compared to the characteristic time of the induced star formation.

In this work, we extended the study of the metallicity distribution of circumgalactic clouds to the extremely low metallicity range, $[\rm{X/H}]<-1.7$ \linebreak ($[\rm{Fe/H}]<-2.0$) (Nu{\~n}ez et al., 2022). The extremely low metallicity component was revealed in low-density clouds (partial Lyman limit systems, pLLSs) with neutral hydrogen linear densities in the range of \mbox{$16.1<\log N_{\rm HI}<17.2$} (see, for example, \mbox{Figs.~1--3} in Acharova et al. (2022), where more than two maxima are clearly visible in the histograms), and at any redshift.
To illustrate the relationship between circumgalactic clouds and GCs in the entire observed metallicity range, we present histograms in Figs.\;1 and\;2. The statistics of the distributions shown in these figures are presented in Table\;1.
Figure~1a shows the distribution of the silicon abundance in dense clouds (damped Lyman limit systems, the linear density of neutral hydrogen for them  is in the range of \mbox{$\log N_{\rm HI} > 20.3$)} for redshifts \mbox{$5<z<2$} (Nu{\~n}ez et al., 2022). In Nu{\~n}ez et al. (2022), only circumgalactic clouds with extremely low metallicity, \mbox{$[\rm{X/H}]<-1.7$ ($\rm [Fe/H]<-2.0$)}, were studied. 
Figure\;1b shows the metallicity distribution of circumgalactic clouds (pLLSs) (Lehner et al., 2019). Metallicity was determined primarily from oxygen ions (a combination of O\,II, O\,III, O\,IV, and occasionally O\,I and O\,VI), magnesium, and/or silicon.
Figure\;1c shows the distribution constructed using the lowest metallicity values of GCs from 28\;galaxies (see Table\;2 in Beasley et al., 2019). To clarify, GCs of different metallicities occur within a galaxy, and to construct the histogram, we used a single, lowest-metallicity cluster from each galaxy to ensure that it formed from circumgalactic gas.

Figure\;2a shows the histogram of the $[\rm{X/H}]$ distribution for the GCs of the Galactic disk from the study by Dias et al. (2016), the metallicity of which has been determined from the magnesium abundance; Fig.\;2b is for the pLLS circumgalactic clouds from the work by Lehner et al. (2019) (the same diagram as in Fig.\;1b, to facilitate comparison with the GC distributions); Fig.\;2c is for the GCs of the Galactic halo from Pritzl et al. (2005).

% Fig 1
 \begin{figure}[ht]
 %\onelinecaptionstrue \captionstyle{normal} %\setcaptionmargin{5mm}
 \includegraphics[width=0.95\columnwidth]{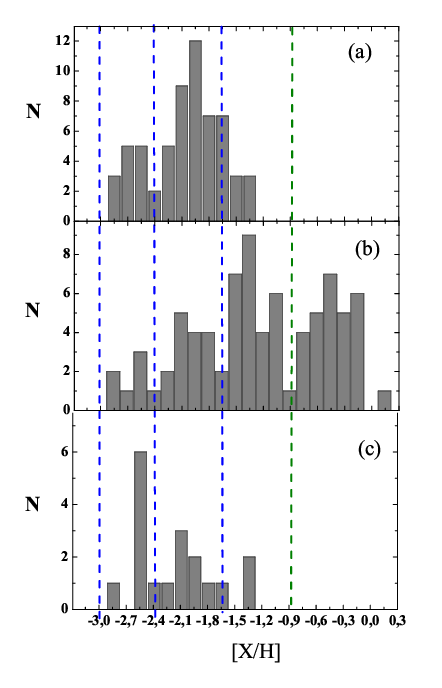}
 \caption{Histograms of the metallicity distribution of the following objects:
panels\;(a) and (b) show high- and low-density circumgalactic clouds from Nu{\~n}ez et al. (2022) and Lehner et al. (2019), respectively; panel (c) shows GCs from 28\;galaxies selected based on their minimum metallicity values (Beasley et al., 2019). The vertical dashed lines pass through the minima in the metallicity distribution of pLLS circumgalactic clouds (b) to facilitate comparison of all the histograms shown in the figure.}
 \label{fig1}
 \end{figure}

 % Fig 2
 \begin{figure}[ht]
 %\onelinecaptionstrue \captionstyle{normal} %\setcaptionmargin{5mm} \vspace{-0.5mm}
  \includegraphics[width=0.95\columnwidth]{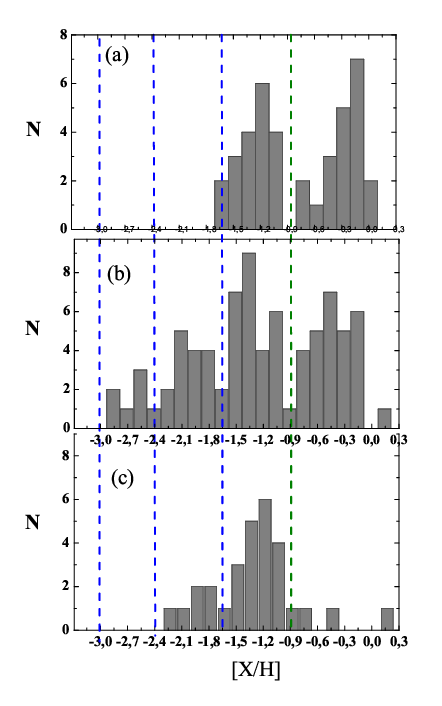}
 \caption{The same as in Fig.\;1, but for the following objects: in panel\;(a)~globular clusters located in the disk of our Galaxy (Dias et al., 2016); (b)~low-density circumgalactic clouds from Lehner et al. (2019) (the same diagram as in Fig.\;1b is repeated here for easier comparison); (c)~GCs located in the halo including its outermost parts (Pritzl et al., 2005). It is clear that the maxima in different samples coincide within the standard deviation.}
 \label{fig2}
 \end{figure}

Based on the analysis of the histograms in Figs.\;1 and\;2, several ranges of metallicity can be identified: \begin{list}{}{
\setlength\leftmargin{8mm} \setlength\topsep{0mm}
\setlength\parsep{0mm} \setlength\itemsep{0.5mm} }
    \item [1)] extremely low  \mbox{($[\rm{X/H}]<-1.7$)},
    \item [2)] low \mbox{($-1.7 \le[\rm{X/H}]<-0.9$)}, 
    \item  [3)] high  \mbox{($ [\rm{X/H}]\ge-0.9$)}.
    \end{list} Moreover, the extremely low-metallicity region, in turn, is apparently divided into two subregions with a minimum at approximately  \mbox{$ [\rm{X/H}] \simeq -2.3$}. In our presentation, we also decided to adhere to the terminology used in Lehner et al. (2019) and {Nu{\~n}ez} et al. (2022). For the extremely low-metallicity range, we will specify the subgroup by indicating the metallicity. 
 \begin{table*}[]
\caption{Average values of $[\rm{X/H}]$, standard deviations \mbox{$\sigma([\rm{X/H}])$}, and the number $N$ of GCs and circumgalactic clouds in different metallicity ranges in accordance with the data by 
Pritzl et al. (2005), Dias et al. (2016), Beasley et al. (2019),  Lehner et al. (2019), and {Nu{\~n}ez} et al. (2022)} \medskip
\begin{tabular}{l|l|c|c|c|c|c|c}
\hline
\multicolumn{1}{c|}{\multirow{2}{*}{Object}}& \multicolumn{1}{c|}{\multirow{2}{*}{Reference}}&
$\langle [\rm{X/H}] \rangle$& $\sigma([\rm{X/H}])$& $N$ &$\langle [\rm{X/H}] \rangle$& $\sigma([\rm{X/H}])$& $N$\\  \cline{3-8}
&&\multicolumn{3}{c|}{$[\rm{X/H}] <-2.3$} &\multicolumn{3}{c}{$ -2.3 \le [\rm{X/H}]<-1.7 $} \\  \hline
pLLSs &Lehner et al. (2019)   & $-$2.63 & 0.17 & 7 & $-$1.98 & 0.16 & 16 \\ 
DLAs &{Nu{\~n}ez} et al. (2022)   & $-$2.66 & 0.29 & 33 & $-$2.11 & 0.12 & 28 \\ 
GC& Beasley et al. (2019)   & $-$2.5 & 0.12 & 8 & $-$2.0 & 0.16 & 7 \\ 
GC &Pritzl et al. (2005)  & -- & -- & -- & $-$1.93 & 0.10 & 6 \\ 
GC &Dias et al. (2016)  & -- & -- & -- & -- & -- & -- \\ \hline
\multicolumn{2}{c|}{}&\multicolumn{3}{c|}{$  -1.7 \le [\rm{X/H}] < -0.9$} &\multicolumn{3}{c}{$ [\rm{X/H}]\ge -0.9$} \\  \hline
pLLSs &Lehner et al. (2019)   & $-$1.31 & 0.19 & 28 & $-$0.42 & 0.22 & 28 \\ 
DLAs &{Nu{\~n}ez} et al. (2022)   & -- & -- & -- & -- & -- & -- \\ 
GC &Beasley et al. (2019)  & $-$1.42 & 0.18 & 3 & $-$ & $-$ & $-$ \\ 
GC &Pritzl et al. (2005)  & $-$1.25 & 0.18 & 20 & --& -- & -- \\ 
GC& Dias et al. (2016)  & $-$1.28 & 0.20 & 19 & $-$0.28 & $-$0.20 & 20 \\ \hline
\end{tabular}
\label{tab:colucci17}
\end{table*}
\renewcommand{\baselinestretch}{1}

Let us now analyze the statistics of the distributions shown in Figs.\;1 and\;2 constructed in accordance with the observations by Pritzl et al. (2005), Dias et al. (2016), Beasley et al. (2019), Lehner et al. (2019), and {Nu{\~n}ez} et al. (2022) given in Table\;1 for different metallicity ranges. Note that cloud subgroups of all four metallicity ranges are presented only in Lehner et al. (2019). Moreover, the average values in different samples of objects within each metallicity range coincide within the standard deviation. Therefore, we will consider the sample by Lehner et al. (2019) to closely reflect the realistic picture of the distribution of circumgalactic clouds and GCs by metallicity. As can be seen from the statistical analysis of the data from Lehner et al. (2019), the number of clouds in the metallicity ranges under consideration is \mbox{$7: 16: 28: 28$}. However, in the range of  \mbox{$[\rm{X/H}] <-2.3$}, Lehner et al. (2019) were able to determine only the upper metallicity limit for 13\;more clouds. They were not included in our statistical analysis, but taking them into account brings the number of clouds to \mbox{$20: 16: 28: 28$}. Thus, in the extremely low-metallicity group \mbox{($[\rm{X/H}] < -1.7$)}, the clouds are distributed almost equally among the subgroups. The number of circumgalactic clouds in the low- and high-metallicity subgroups is the same (Acharova et al., 2022). It will be shown below (see Section\;4) that the exact value of the cloud proportion is not critical. For example, underestimating or overestimating by a factor of two the number of clouds in any metallicity range considered will lead to a corresponding underestimation or overestimation by a factor of two of the mass of metals produced by supernovae. However, the conclusions about the number of supernovae will remain unchanged. Note that in this case, the set of the maximum masses of stars capable of exploding as supernovae, obtained for average metallicities within each range, will change according to the modified metal production proportion (see Section\;4).

Comparing the histograms of the distribution of the number of objects from the metallicity of two subgroups of circumgalactic clouds at \mbox{$[\rm{X/H}]\le -1.7$} and the lowest metallicity of GCs in nearby dwarf galaxies from Beasley et al. (2019) (see Fig.\;1), we come to the conclusion that GCs can form in clouds with \mbox{$[\rm{X/H}] \ge -3.0$}. In giant galaxies MW, M\,31, Cen\,A, M\,87, as can be seen from Fig.\;7 by Beasley et al. (2019), GCs with \mbox{$ \rm [\rm{X/H}] \le -2.3$} have not been preserved and with \mbox{$-2.3 < [\rm{X/H}] \le -1.7$}, only a few GCs remained. In any case, the occurrence of GCs at \mbox{$[\rm{X/H}] < -1.7$} in giant galaxies is significantly lower than the occurrence of clouds: \mbox{$20: 16: 28: 28$}. The reason for this remains to be clarified. Since if we adhere to the hierarchical theory of galaxy formation, then, for example, about 50\% of GCs in our Galaxy have been formed in dwarf galaxies and have been subsequently accreted (Kruijssen et al., 2019; Massari et al., 2019; Forbes, 2020). It is interesting to note that all known ultra-faint dwarf galaxies began to form at metallicities smaller than $-$3.0\,dex (Fu et al., 2023), that is, the limit metallicity of UFDs lies to the left of all the histograms shown in Figs.\;1 and\;2. Let us consider separately the properties of UFDs and GCs as representatives of objects formed in the same epoch.

\section{ANALYSIS OF THE OBSERVED DATA}
\subsection{Ultrafaint Dwarf Galaxies: Properties and Formation}

Ultrafaint dwarf galaxies are the oldest stellar objects. Their star formation ended 12\;Gyr ago (Brown et al., 2014) and occurred under conditions that have no analogue in the modern era: rarefied gas was likely almost completely fragmented into stars due to the gravitational influence of the dark matter. In support of this conclusion, we present the following estimates. The mass of the stellar component of UFDs is \mbox{$10^2$--$10^3\,M_{\odot}$} with a characteristic size of about $100$\,pc and the virial masses from $10^7\,M_{\odot}$ to $10^9\,M_{\odot}$ (Fu et al., 2023). According to the Salpeter IMF (1955), the stars with masses between $0.1\,M_{\odot}$ and $0.8\,M_{\odot}$, that survived in UFDs, account for about 56\% of the mass of the original star cluster, i.e., the initial mass of the stars was only twice as large.
Let us make an upper bound: assume that the mass of gas before star formation in UFDs was an order of magnitude greater than the stellar mass it formed. We get a surface gas density no higher than $10\,M_{\odot}\,$pc$^{-2}$. On the other hand, modern observations of dwarf galaxies and the outskirts of giant galaxies show that at surface gas densities below $10\,M_{\odot}\,$pc$^{-2}$, the efficiency of star formation is so low that the characteristic time of gas exhaustion is $10^{10}$\,years and even $10^{11}$\,years (Bigiel et al., 2008). In this case, open clusters of the Pleiades type are typically formed (Bigiel et al., 2008), which do not contain stars with masses greater than $8\,M_{\odot}$ capable of becoming supernovae. In UFDs, the observed density of stars indicates that Pleiades-like clusters were formed (Simon, 2019; Fu et al., 2023), but the main episode of star formation lasted about 100\;Myr (Jones et al., 2025). Such a sharp difference in the time of star formation can be explained, if we take into account that the star formation rate is affected not only by the gas density, but also by the total surface density, which has been first substantiated in the paper by Talbot and Arnett (1975). In their study, Portinari and Chiosi (1999) compared the influence of different star formation laws on the results of modeling the evolution of the Galactic disk including its chemical evolution. As a result, it was demonstrated that the value of the proportionality factor $\nu$ used in Schmidt's law (Schmidt, 1959) \mbox{$\psi=\nu\mu_g^k$} ($\mu_g$ is the surface density of gas, $k$ is the exponent), implicitly contains a contribution from the total surface density of the matter and, strictly speaking, is a function of the galaxy's radius. Note that only in UFDs, the mass of the dark matter exceeds the mass of the visible baryonic matter by 5--6 orders (Simon, 2019; Fu et al., 2023). In other words, there are reasonable grounds to believe that in UFDs, a short-term burst of star formation occurs under conditions of the low gas density under the influence of the dark matter. During the star-formation burst, an almost complete, up to 100\%, exhaustion of star-forming gas is usually observed (Kennicutt and Evans, 2012). 

Detailed modeling of the UFD evolution under the influence of a dominant dark matter contribution is not yet available. However, given the above, we do not consider UFDs to be objects capable of making a significant contribution to the enrichment of the extremely low-metallicity component of circumgalactic clouds.

\subsection{Globular Clusters: Formation and Properties}

The beginning of GC formation coincided with the formation of UFDs: approximately 13\;Gyr ago. The properties of these two types of objects are diametrically opposed: GCs are characterized by very high stellar densities and do not contain the dark matter. There is an extensive amount of literature on the formation of GCs (e.g., Kruijssen, 2015; Li et al., 2017; Pfeffer et al., 2018; Forbes, 2020). The main conclusion of these studies is that for GCs to form, the gas density and star-formation rate must be sufficiently high.

The metallicity distribution of the lowest-metalli\-ci\-ty GCs from 28\;galaxies according to Table\;2 of Beasley et al. (2019) coincides with the metallicity distribution of the extremely low-metallicity cloud subgroup with \mbox{$[\rm{X/H}] < -1.7$} (see Fig.\;1 and Table\;1 of this paper). Since the sample is represented mainly by dwarf galaxies, it can be seen that GCs began to form in extremely low-metallicity gas, in two of its subranges: \mbox{$[\rm{X/H}] < -2.3$} and \mbox{$-2.3 \le [\rm{X/H}] < -1.7$}. Some dwarf galaxies, such as Sextant~A, Pegasus~dIrr, and WLM, may contain only one GC. Others may contain several GCs. For example, in NGC 147 and NGC 6822, the GCs fall into both subgroups of the extremely low-metallicity range, \mbox{$[\rm{X/H}] \le -1.7$}. This distribution may be a portrait of the metallicity of the medium from which GCs formed, that is, it reflects the degree of metal mixing within the gaseous cloud, from which the dwarf spheroidal galaxy formed. The metallicity distribution within a dwarf galaxy could be highly heterogeneous even at its early stages. The proximity of two star-forming centers with extremely high gas densities occurs in shells formed by the expanding gas of exploded supernovae (Vasiliev et al., 2017; Barnes et al., 2023). However, a detailed discussion of the latter issue is beyond the scope of this paper.

The GCs belonging to our Galaxy and M\,31 are located predominantly in the range of \mbox{$[\rm{Fe/H}] \ge-1.7$} (Carretta et al., 2010; Dias et al., 2016; Wang et~al., 2021). Only a few of them fall into the range of \mbox{$-2.3 \le [\rm{Fe/H}] < -1.7$}. There is a discrepancy with the metallicity distribution of GCs in dwarf galaxies (Beasley et al., 2019). 

Currently, the masses of most GCs in the Milky Way are in the range of \mbox{$10^5$--$10^6\,M_{\odot}$} (Gratton et al., 2019). However, according to a number of studies (e.g., Winter and Clarke, 2023 and references therein), GCs have lost up to 90\% of their mass by now during interactions with our Galaxy. This means that the mass of many of them could initially have been of the order of $10^7\,M_{\odot}$.
In the paper by Kruijssen et al. (2019), it is concluded that the metallicity limit for GCs at the level of \mbox{$[\rm{X/H}]=-2.5$} is a result of the observed ``mass--metallicity'' relation of the host galaxies (the smaller the galaxy, the lower the metallicity of the stars inhabiting it). It should be noted that the conclusion of Kruijssen et al. (2019) was made by extrapolation and not from direct observations. In contrast, one can point to the study of Beasley et~al. (2019). Sometimes, a single GC is observed in galaxies formed at the initial stage of galaxy formation and its metallicity does not correlate with the average mass of stars in the modern galaxy. It is clear that galaxies with higher masses are capable of forming a larger number of GCs involving the matter enriched in the first GCs too. The average metallicity of several generations of GCs will be higher than that of the first generation of GCs. Ultra-compact dwarf galaxies and some dwarf spheroidal galaxies of the Local Group (Figs.\;9 and 11 in the paper by Chattopadhyay et al., 2012) also do not exhibit the ``mass--metallicity'' relation. 

To estimate the properties of globular clusters in different metallicity ranges, we turn to calculations of the chemical evolution of circumgalactic clouds.

 \section{CALCULATING THE NUMBER OF CORE-COLLAPSE SUPERNOVAE STARS THAT FORMED THE OBSERVED METALLICITY OF CLOUDS}

Calculating the number of core-collapse supernovae (i.e., those originating from stars with masses greater than 8\,$M_{\odot}$) capable of enriching clouds from the average value in one metallicity range to the average value in a neighboring metallicity range, as will be shown in this section, can provide information about the mass of the host star clusters and about the dependence of the maximum presupernova mass on metallicity.

We begin our discussion from the moment when, as a result of the first supernova explosions, the circumgalactic clouds have already reached an extremely low level of metallicity. The only stellar systems that have the metallicities \mbox{$[\rm{X/H}] < -3$}~ are ultrafaint dwarf galaxies (UFDs). GCs were formed in the range of \mbox{$-3 \le [\rm{X/H}]<-2.3$}, and the mechanism for such a difference in metallicity between UFDs and GCs remains beyond the scope of this study.

Let us qualitatively discuss the process of enrichment of circumgalactic clouds with heavy elements. First, globular clusters formed in extremely low-metallicity gas (\mbox{$-3 \le [\rm{X/H}]<-2.3$}). Within a short period of time, core-collapse supernovae (which, according to Salpeter's initial mass function (Salpeter, 1955), account for approximately 10\% of the globular cluster mass) explode. As a result, the gas surrounding the globular clusters is compressed by a shock wave and enriched in heavy chemical elements. Then, a certain percentage of the mass of this gas, 10\% and higher (Kennicutt and Evans, 2012), becomes the matter of globular clusters with the metallicity \mbox{$-2.3 \le [\rm{X/H}]<-1.7$}. Star formation processes in the shells of bubbles surrounding massive star clusters are still observed today (Watkins et al., 2023 and references therein). Further, as a result of a similar process, two more generations of globular clusters are formed at the metallicities \mbox{$-1.7 \le [\rm{X/H}]<-0.9$} and \mbox{$[\rm{X/H}] \ge -0.9$}. The metallicity distribution of clouds and globular clusters with clearly definite maxima indicates that gas enrichment occurred sequentially from extremely low to high metallicity.
This, in turn, suggests that similar processes with the coherent explosions of hundreds of thousands of supernovae occurred in different parts of the Universe.

Next, let us analyze the enrichment process of a single cloud. To more clearly present our reasoning, we assume that clouds are currently equally distributed among subgroups in different metallicity ranges. An illustration of the sequential enrichment of a circumgalactic cloud is shown in Fig.\;3. We assume that initially, all the gas involved in the formation of GCs had the metallicity  \mbox{$-3 \le [\rm{X/H}] < -2.3$}. Globular clusters formed in this gas enriched $3/4$ of the cloud's mass to \mbox{$-2.3 \le [\rm{X/H}] < -1.7$}. Then supernovae exploded in clouds with the metallicity \mbox{$-2.3 \le [\rm{X/H}] < -1.7$} enriched $2/3$ of the mass of this gas to \mbox{$-1.7 \le [\rm{X/H}] < -0.9$}. Finally, supernovae of low-metallicity clouds enriched half the mass and created a high-metallicity subgroup with \mbox{$[\rm{X/H}] \ge  -0.9$}. The clouds retained the metallicity that remained from coherent supernova explosions in the GC.

 % Fig 3
 \begin{figure*}[ht] 
 %\onelinecaptionstrue \captionstyle{normal} %\setcaptionmargin{5mm}
 \includegraphics[width=0.85\textwidth]{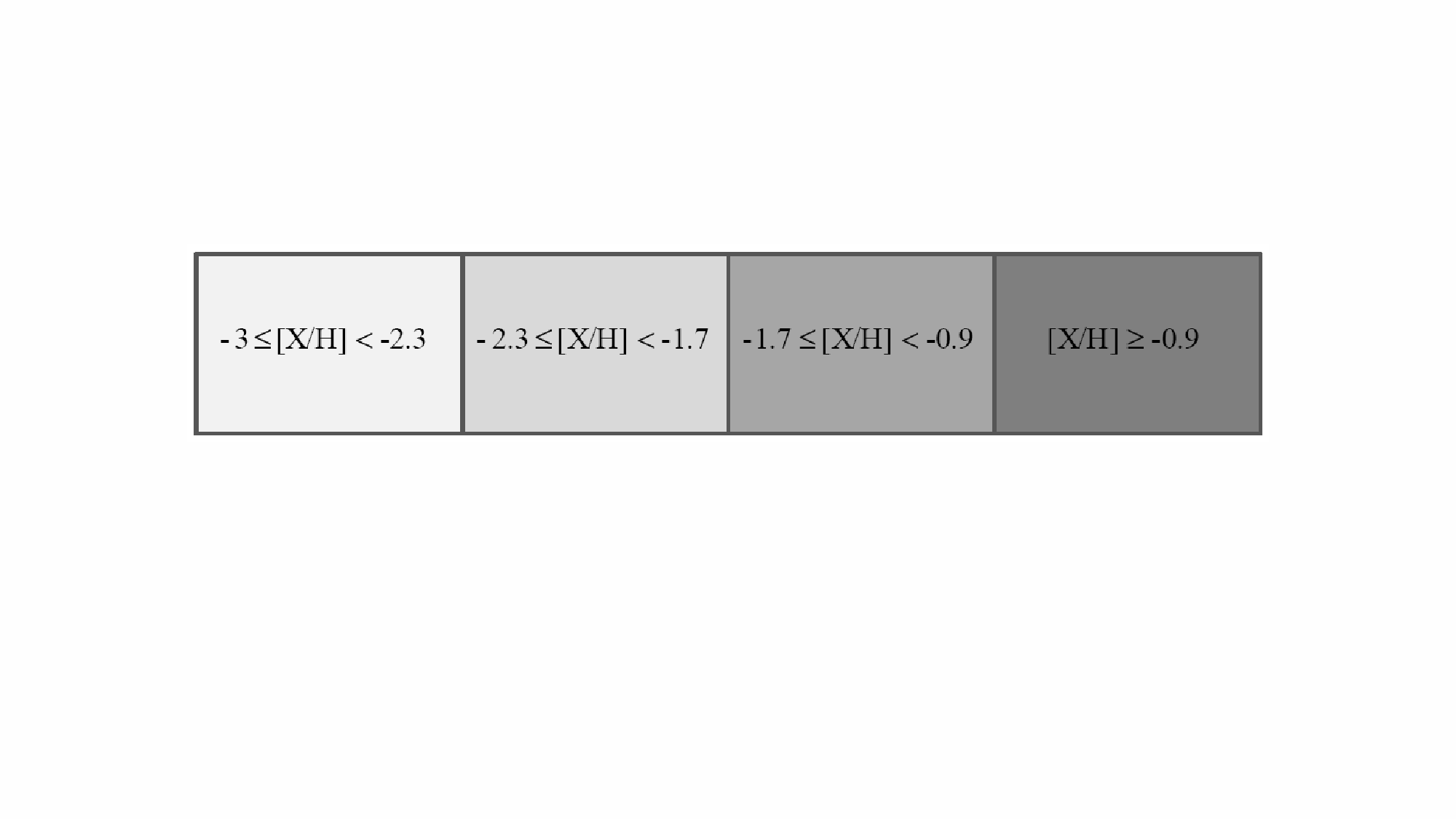}
 \vspace{-3cm}
 \caption{Diagram explaining the enrichment of the initial gaseous cloud assuming that there is now an equally probable distribution of the number of circumgalactic clouds over metallicity ranges. Initially, the entire cloud had the metallicity \mbox{$[\rm{X/H}] <-2.3$}. Then $3/4$ of the cloud mass was enriched to \mbox{$-2.3 \le [\rm{X/H}]<-1.7$}, from which $2/3$, in turn, to \mbox{$-1.7 \le [\rm{X/H}] < -0.9$}. And finally, half the mass with the metallicity \mbox{$-1.7 \le [\rm{X/H}] < -0.9$} was enriched to  \mbox{$[\rm{X/H}] \ge -0.9$}.}
 \label{fig3}
 \end{figure*}
 
\begin{table*}
%\setcaptionmargin{0mm} \onelinecaptionstrue \captionstyle{normal} \setlength{\tabcolsep}{3pt}
\caption{Calculations of oxygen enrichment in circumgalactic clouds } \medskip
\begin{tabular}{c|c|c|c} \hline 
Average metallicity&  $\langle [\rm{X/H}] \rangle =-2.63\!\div\! -1.98$ &   $\langle  [\rm{X/H}] \rangle =-1.98\!\div\!-1.31$ & $\langle [\rm{X/H}] \rangle=-1.31\!\div\!-0.42$ \\
 \hline
Oxygen mass, $M_{\odot}$  & $10^{2.52}$  & $10^{3.01}$  &$10^{3.65}$ \\
\hline
 \end{tabular}
\label{tab:oxygen}
\end{table*} 

We will consider the enrichment of a gaseous cloud with a mass of, for example, $10^{8}\,M_{\odot}$. As shown in Nagakura et al. (2009), $10^{8}\,M_{\odot}$ is the critical mass, starting from which supernova-induced star-formation bursts can occur. In the study by Lehner et al. (2019), the clouds occupy the entire analyzed metallicity range: from extremely low-metallicity to high-metallicity. Therefore, we use the average metallicity values of all four cloud subgroups from this study by Lehner et al. (2019). In the other studies shown in Figs.\;1 and\;2, the considered metallicity range is narrow, and only two maxima appear in the distribution histogram.

The metallicity of the circumgalactic clouds in Lehner et al. (2019) was determined mainly from the abundances of oxygen and magnesium ions. Let us estimate the oxygen mass required to enrich a cloud from an average metallicity of $-2.63$ for the range of \mbox{$[\rm{X/H}] < -2.3$} to an average metallicity of $-1.98$ for the range of \mbox{$-2.3 \le [\rm{X/H}] < -1.7$} (see Table\;1).
By definition, the mass fraction of oxygen is calculated using the formula:  
\begin{eqnarray} 
\nonumber 
Z_{\rm O} = 10^{\log Z_{\rm O_{\odot}} + [\rm{X/H]}},  \nonumber 
% \label{brightness}
\end{eqnarray}
where  \mbox{$\log Z_{\rm O_{\odot}}=-3.27$}  (Asplund et~al., 2009).

According to the average metallicity values: \linebreak \mbox{$\langle [\rm{X/H}] \rangle=-2.63$} and  \mbox{$\langle [\rm{X/H}] \rangle=-1.98$}, we have two values of the mass fraction of oxygen:  {$Z_{\rm O}=10^{-2.63-3.27}\!=\!10^{-5.90}$} and \mbox{$Z_{\rm O}\!=\!10^{-1.98-3.27}\!=\!10^{-5.25}$}\!\!.

Taking into account the above reasoning that the first generation of GC supernovae enriched $75\%$ of the circumgalactic cloud, it can be argued that the enriched part of the cloud acquired the following mass of oxygen:  \begin{equation}
\begin{array}{rcl}
\!\!M_{\rm O}\!=\!0.75\!\times\!10^{8}(10^{-5.25}\!-\!10^{-5.90})M_{\odot}\!\approx\!10^{2.52}M_{\odot}. \nonumber 
% \label{brightness}
\end{array}
\end{equation}
It is clear that the average values of $[\rm{O/H}]$ are determined based on the available cloud sample and may differ from the true average value due to the limitations of the sample we use.

Next, we will estimate the mass of oxygen required to enrich half the mass of the original cloud from the state with the average value \mbox{$\langle [\rm{X/H}] \rangle=-1.98$} to a state with the average value \mbox{$\langle [\rm{X/H}] \rangle=-1.31$}. It is: \begin{equation}
\begin{array}{rcl}
 \!M_{\rm O}\!=\!0.5\!\times\! 10^{8}(10^{-4.58}\!-\!10^{-5.25})\,M_{\odot}\! \approx\!10^{3.01}M_{\odot}. \nonumber 
% \label{brightness}
\end{array}
\end{equation}

Finally, we estimate the mass of oxygen required to enrich a quarter of the mass of the original cloud from the state with the average value \mbox{$\langle [\rm{X/H}] \rangle=-1.31$} to the state with the average value  \mbox{$\langle [\rm{X/H}] \rangle = -0.42$}: \begin{equation}
\begin{array}{rcl}
\!\!M_{\rm O}\!=\!0.25\!\times\!10^{8}(10^{-3.59}\!-\!10^{-4.58})M_{\odot}\!\approx\! 10^{3.65}M_{\odot}.
\nonumber 
% \label{brightness}
\end{array}
\end{equation}

That is, there is a clear trend towards an increase in the required oxygen mass to enrich an increasingly smaller portion of the initial gaseous cloud, in which four generations of GCs formed sequentially.

As can be seen from the estimates obtained, changing the cloud mass, i.e., using a different value instead of $10^{8}\,M_{\odot}$, will lead to a change in the oxygen mass by the same proportion in each of the three groups, but will not lead to a change in the proportion. The masses given in Table\;2 are calculated for a given cloud mass of $10^{8}\,M_{\odot}$.

As can be seen from the analysis of Table\;2, with increasing metallicity of a part of the original cloud, the proportion of the produced oxygen increases by a factor of\;3 and 4.4. At the same time, there is no dependence of the GC mass on metallicity, which is consistent with the observed data (Harris, 2010; Krause et~al., 2016). Consequently, there is no reason to expect a similar increase in the number of type\;II supernovae originating from single stars. Let us consider whether type\;Ib/c supernovae originating from binary stars can make the necessary contribution. There are no observational constraints on the proportion of different types of core-collapse supernovae in GCs. A wide scatter of the ratio of the number of type\;II supernovae to the number of type\;Ib/c supernovae is observed for galaxies (Heger et~al., 2003)---\mbox{from 1.2} \mbox{to 16}. However, stellar densities in modern star-forming centers are three orders lower than in GCs. As discussed, for example, in Gratton et al. (2019), under conditions of high stellar density, close binary star systems are effectively disrupted. Binary stars account for less than 10\% of the total GCs, and this ratio is independent of metallicity. Therefore, there is no obvious reason to expect a 3--4-fold increase in the supernova formation rate due to the contribution of type\;Ib/c supernovae. The only possible increase in the oxygen mass produced during the core-collapse supernova bursts in GCs is an increase in the maximum mass of stars capable of exploding as supernovae with increasing metallicity of the source gas.

As can be seen from Fig.\;4, which shows the average mass of oxygen per supernova obtained taking into account the initial Salpeter mass function, according to nucleosynthesis calculations (Tsujimoto et~al., 1995; Nomoto et~al., 2006), to increase the mass of the produced oxygen by a factor of three, it is sufficient to increase the maximum mass of stars in the cluster capable of exploding as supernovae, for example, from $11.5\,M_{\odot}$ to $13\,M_{\odot}$ (Nomoto et~al., 2006). With further increase in metallicity, in order to increase the mass of the produced oxygen by a factor of 4.4, the maximum mass of the presupernova must be increased to $20\,M_{\odot}$. In this case, the average mass of oxygen per supernova varies from $0.012\,M_{\odot}$ to $0.036\,M_{\odot}$ and $0.16\,M_{\odot}$, respectively  (Nomoto et~al., 2006). As can be seen from Fig.\;4, the maximum masses of presupernova stars for different metallicity ranges cannot be unambiguously indicated from the information on the proportion of the produced oxygen alone: a given proportion corresponds to an arbitrary choice of other mass values, for example, $12.5\,M_{\odot}$, $15\,M_{\odot}$, and $25\,M_{\odot}$ (while the average mass of oxygen per supernova according to Nomoto et~al. (2006) varies from $0.025\,M_{\odot}$ to $0.075\,M_{\odot}$, and $0.33\,M_{\odot}$, respectively). The limit on the maximum presupernova mass in this trio of values may be a theoretical limit obtained from stellar evolution studies, sensitive to model parameters, and being at $25\,M_{\odot}$ (Fryer and Kalogera, 2001; Heger et~al., 2003).  A similar value, $23\,M_{\odot}$, is obtained by modeling the chemical evolution of the Galaxy (Acharova et~al., 2013). Given this limitation, we find that the set of possible values for the maximum presupernova mass for different metallicity ranges can be fixed at the top. The calculations by Tsujimoto et~al. (1995), shown in Fig.\;4, demonstrate a steeper dependence of the average oxygen mass per supernova on the maximum presupernova mass. Therefore, the range of presupernova masses for different metallicities according to Tsujimoto et~al. (1995) will be slightly narrower than that for the data from Nomoto et~al. (2006). Similar reasoning applies to magnesium production.

The dependence of the maximum presupernova mass on metallicity has already been considered in the literature. The physical mechanisms responsible for its variation have been identified, although, the value of this mass remains model-dependent. For example, Heger et al. (2003) noted that at low metallicities, stars with masses above $25\,M_{\odot}$ do not explode as supernovae, but rather undergo direct collapse of the stellar matter into a black hole. With increasing metallicity, the stellar wind increases, which affects the mass of the star and, thereby, indirectly affects the mass of the stellar core leading to a state, in which a supernova explosion is probable. This conclusion is indirectly confirmed by the fact that gamma-ray bursts are known to be more often observed in low-metallicity systems, a fact that is theoretically substantiated in Woosley and Heger (2006). In these cases, as a rule, the ejection of the enriched supernova shell into the surrounding space does not occur (Fryer et al., 1999). The study presented in this paper represents an independent method for determining the dependence of the maximum presupernova mass on metallicity.

 % Fig 4
 \begin{figure}[ht]
 %\onelinecaptionstrue \captionstyle{normal} \setcaptionmargin{3mm} \vspace{2mm}
 \includegraphics[width=0.43\textwidth]{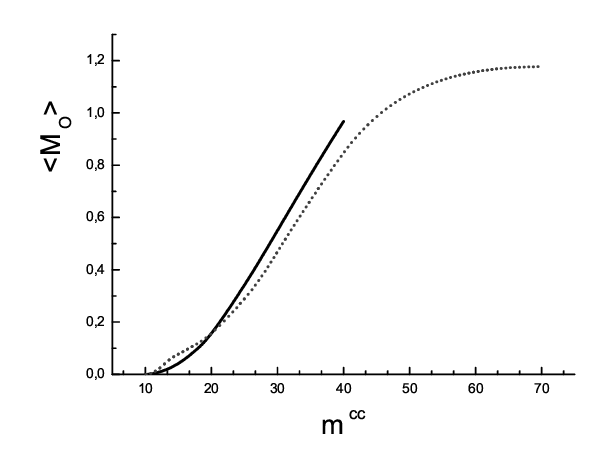}
 \caption{Average mass of oxygen produced per one supernova as a function of the maximum mass of stars in the star cluster that can explode as supernovae derived from the nucleosynthesis data: Tsujimoto et al. (1995)---the solid line, Nomoto et al. (2006)---the dashed line.}
 \label{fig4}
 \end{figure}

Now we have enough information to begin studying the star formation in circumgalactic clouds. Since there is no dependence of the GC mass on metallicity (Harris, 2010; Krause et~al., 2016), in the first step we will assume that the GC masses in different metallicity ranges are the same. Therefore, they contain an equal number of stars with masses greater than $8\,M_{\odot}$, and this means that the number of supernovae in different metallicity ranges is the same. Suppose that in the metallicity range  \mbox{from $\langle [\rm{X/H}] \rangle =-1.98$}   \mbox{to $\langle [\rm{X/H}] \rangle=-1.31$}, the maximum presupernova mass is equal to $15\,M_{\odot}$. Then the required number of supernovae:  \mbox{$10^{3.01}\,M_{\odot}/ 0.075\,M_{\odot}= 10^{4.13}$}. 

The average supernova mass $\langle\rm  Mcc \rangle$ assuming that the maximum presupernova mass is $15\,M_{\odot}$ is determined by the formula:
$$\langle {\rm Mcc} \rangle = {{\!\!\int_{8}^{15}\!\! m \varphi(m) dm}\over{\!\!\int_{8}^{15}\!\! \varphi(m) dm}} \approx 11\,(M_{\odot}),$$ 
where $m$ is the mass of stars, $\varphi(m)$ is the Salpeter's initial mass function.

Thus, the mass of supernovae in a star cluster:
$10^{4.13} \times 11=10^{5.18}\,(M_{\odot}).$
The fraction of the supernova masses in the star cluster in this case is $$\int_{8}^{15}\!\! m \varphi(m) dm \approx 0.05.$$ 
This means that supernovae with a total mass of $10^{5.18}\,M_{\odot}$ are formed in GCs with a mass of \linebreak \mbox{$3 \!\times\!10^{6}\,M_{\odot}.$}

Therefore, from the above calculations, it follows that in order to enrich a cloud of an initial mass of $10^{8}\,M_{\odot}$ with oxygen, so that at the end of the process of sequential enrichment it consists of four regions of metallicity equal in mass studied in the paper: \mbox{$[\rm{X/H]} <-2.3$},  \mbox{$-2.3 \le [\rm{X/H}]<-1.7$},  \mbox{$-1.7 \le [\rm{X/H}] < -0.9$},  and \mbox{$[\rm{X/H}] \ge -0.9$}, it is sufficient that in each region one or more GCs with a total mass of \mbox{$3 \times 10^{6}\,M_{\odot}$} are formed.

Calculations show that if the mass of a GC is smaller than a critical value of $2 \times 10^{5}\,M_{\odot}$, it will be destroyed in a time of about 10\;Gyr when moving in the gravitational field of the Galaxy (Kruijssen, 2015; Gratton et al., 2019).
There are indications that the initial masses of the GCs that have survived to the present day in our Galaxy were an order of magnitude greater than modern ones  (Harris, 2010; Kruijssen, 2015; Gratton et al., 2019). That is, the masses of the most massive GCs could be of the order of $10^{7}\,M_{\odot}$. This means that the masses of the clouds that formed them were 2--3\;times greater than $10^{8}\,M_{\odot}$ (this value was found in Nagakura et al. (2009) as the critical valuem starting from which star formation bursts can be induced by supernovae). If fifteen\;GCs with a mass of \mbox{$2 \times 10^{5}\,M_{\odot}$} (a total mass of \mbox{$3 \times 10^{6}\,M_{\odot}$)}, then they were mostly destroyed while moving in the gravitational field of our Galaxy but could have been preserved, as we see, in dwarf galaxies (Beasley et al., 2019 and references therein).

From the above, a logical question arises: why are GCs observed in the halo of the Galaxy, as seen from Fig.\;2c, with metallicities in the range of \mbox{$-2.3 \le [\rm{X/H}]<-1.7$}, but not at \mbox{$\rm [\rm{X/H}]<-2.3$}? This issue remains open. However, there are no apparent limitations to the assumption that in extremely low-metallicity regions, star formation conditions were more favorable for the formation of several GCs, whose combined mass corresponded to the mass of a single GC in the higher-metallicity range. It can be noted that GCs aged 10--12\;Gyr located in dwarf galaxies and not experiencing significant disruption have masses close to the critical mass: \mbox{$2 \times 10^{5}\,M_{\odot}$} (Krause et al., 2016). Therefore, GCs, accreted along with their parent dwarf galaxies approximately 10\;Gyr ago, most likely destroyed. 

Suppose that the average metallicity of clouds in one of the subgroups was shifted upward or downward by 0.3\;dex (due to sample limitations or an error in the abundance determination method). This would change the oxygen production by a factor of $10^{0.3}$ and affect the set of values for the maximum masses of stars capable of exploding as supernovae. However, the conclusions about the number of supernovae would remain unchanged, as follows from the above reasoning. 
 
 \section{CONCLUSIONS}
This paper analyzes the conditions for star formation in globular clusters in circumgalactic clouds. The similarity in the metallicity distributions of GCs in the nearby Universe and circumgalactic clouds over a wide range of redshifts, from 0.2 to 5.9, is substantiated in detail. These distributions represent a sequence of four local maxima in the number of circumgalactic clouds and GCs at different metallicities: \mbox{$[\rm{X/H}] \simeq -2.6, -2.0, -1.4, -0.5$}. The successive enrichment of a circumgalactic cloud with a mass of $10^{8}\,M_{\odot}$ from the extremely low metallicity \mbox{$[\rm{X/H}] <-2.3$} has been calculated through the stages of \mbox{$-2.3 \le [\rm{X/H}]<-1.7$}, \mbox{$-1.7 \le [\rm{X/H}] < -0.9$} to the high metallicity \mbox{$[\rm{X/H}] \ge -0.9$}, where the boundaries of the specified ranges coincide with the local minima of the number of objects. It is shown that to reproduce such distributions, it is sufficient that at each stage of enrichment of a portion of the cloud with metals, one or more GCs with a total mass of \mbox{$3 \times 10^{6}\,M_{\odot}$} are formed.

If the initial mass of the most massive GCs observed in the Galaxy was an order of magnitude greater than the modern one, i.e., it reached $10^{7}\,M_{\odot}$ (Kruijssen, 2015; Li et al., 2017; Gratton et al., 2019; Forbes, 2020), then the required mass of their parent circumgalactic cloud is about \mbox{$3 \times 10^{8}\,M_{\odot}$}, otherwise, the observed statistics of the metallicity distribution of clouds and GCs will not be satisfied.

Available observations of the metallicity distribution of GCs in the Milky Way and M\,31 (Beasley et al., 2019 and references in this paper) suggest that in extremely low-metallicity regions (\mbox{$[\rm{X/H}] <-2.3$}) star-formation conditions have been more favorable for the formation of several GCs, the total mass of which corresponded to the mass of a single GC from a more metallic range.

Analysis of the oxygen enrichment of a circumgalactic cloud showed that the maximum mass of stars capable of exploding as supernovae depends on metallicity. A method for detailing this dependence is demonstrated. For the GC metallicities considered in the study, the following maximum presupernova masses are possible: $12.5\,M_{\odot}$, $15\,M_{\odot}$, and $25\,M_{\odot}$ according to calculations of nucleosynthesis in a supernova  (Nomoto et al., 2006). 

\begin{acknowledgments}
The authors are grateful to the reviewer and the Editorial Office of the ``Astrophysical Bulletin'' journal.
\end{acknowledgments}

\section*{FUNDING}
This work was supported by ongoing institutional funding. 
 %No additional grants to carry out or direct this particular research were obtained.
The work by Sh.M.E. was performed as part of the SAO RAS government contract approved by the Ministry of Science and Higher Education of the Russian Federation.

\section*{CONFLICT OF INTEREST}
The authors of this work declare that they have no conflict of interest.

%\bibliographystyle{aspb1}
%\bibliography{Kiyaeva}

%The bibliography without bibtex:

\end{document}